\documentclass[aps,pra,reprint]{revtex4-1}

\usepackage{amsmath,graphicx,physics,tabularx}
\usepackage{lipsum}

\begin{document}

\title{Measurements of the energy distribution of an ultracold rubidium ion beam}

\author{G. ten Haaf}
\author{S.H.W. Wouters}
\author{D.F.J. Nijhof}
\author{P.H.A. Mutsaers}
\author{E.J.D. Vredenbregt}
\email{e.j.d.vredenbregt@tue.nl}
\affiliation{Department of Applied Physics, Eindhoven University of Technology, P.O. Box 513, 5600 MB Eindhoven, the Netherlands}

\date{\today}

\begin{abstract}
The energy distribution of an ultracold rubidium ion beam, which is intended to be used as the source for a focused ion
beam instrument, is measured with a retarding field analyzer. The ions are created from a laser-cooled and compressed
atomic beam by two-step photoionization in which the ionization laser power is enhanced in a build-up cavity. Particle
tracing simulations are performed to ensure the analyzer is able to resolve the distribution. The lowest achieved full
width 50\% energy spread is $\left(0.205\pm0.006\right)$ eV. The energy spread originates from the variation in the
ionization position of the ions which are created inside an extraction electric field. This extraction field is
essential to limit disorder-induced heating which can decrease the ion beam brightness. The ionization position
distribution is limited by a tightly focused excitation laser beam. Energy distributions are measured for various
ionization and excitation laser intensities and compared with calculations based on numerical solutions of the optical
Bloch equations including ionization. A good agreement is found between measurements and calculations.
\end{abstract}


\maketitle

\section{Introduction \label{sec:energy_spread_introduction}}

The ability to modify substrates at the nanometer length scale makes focused ion beams (FIBs) indispensable tools in the
semiconductor industry and nanofabrication research. Important applications are transmission electron microscope (TEM)
sample preparation \cite{Mayer2007}, failure analysis \cite{Donnet2005} and circuit edit \cite{Livengood1999}. The
current industry standard source for FIBs for these nanofabrication purposes is the liquid metal ion source (LMIS). A
LMIS-based FIB reaches a resolution of 5-10 nm with a 1 pA beam of 30 keV ions. Due to chromatic aberrations of the
electrostatic lens system this probe size is for a large part limited by the 4.5 eV (FWHM) LMIS energy spread
\cite{Swanson1980}. Since chromatic aberrations scale with the relative energy spread, this limitation becomes even
more important when a low energy beam is required, for example to lower the ion penetration depth and prevent nanopore
formation or amorphization of crystalline samples \cite{Kolibal2011}.

Ultracold ion sources \cite{McClelland2016} are promising candidates to replace the LMIS in FIBs for nanofabrication
purposes. Recently, for example, a resolution of 2.8 nm (25\%-75\% rise distance, 1.2 pA, 10 keV) was demonstrated with
a focused ion beam equipped with an ultracold cesium ion source \cite{Steele2017}. These ultracold sources are based on the ionization of
laser cooled atoms. The low transverse temperature of the atoms enable ion beam brightnesses of the order of $10^7$
A/m$^2$/sr/eV and energy spreads are expected to be smaller than 1 eV \cite{Knuffman2013,Kime2013,Wouters2014}. Here,
direct measurements of the energy distribution of an ultracold Rb ion beam are presented. These show that the
energy spread can be as low as 0.2 eV and independent of ion beam energy.

The ultracold Rb ion beam is formed in the atomic beam laser cooled ion source (ABLIS). In this source an ultracold
beam of $^{85}$Rb atoms is created by means of magneto-optical compression \cite{tenHaaf2017_atombeam}. After selection of
the desired atomic flux, this beam is photoionized in the crossover of a tightly focused excitation laser beam and an
ionization laser beam whose intensity is enhanced in a build-up cavity \cite{tenHaaf2017_current}. The ions are
immediately extracted by an electric field in order to prevent disorder-induced heating which can lower the ion beam
brightness \cite{tenHaaf2014}. The energy spread in the beam mostly originates from the distribution in the ionization
position within this extraction field. Here, a retarding field analyzer \cite{Simpson1961} is used to measure the
energy distribution. A comparison with expected ionization position distributions \cite{tenHaaf2017_current} is
presented for various experimental parameters.

The remainder of this manuscript is organized as follows. Section \ref{sec:energy_spread_exp_setup} first presents
details of the experimental setup creating the ultracold Rb ion beam and then introduces the retarding field analyzer
that was constructed to measure the energy distribution. Particle tracing simulations, which are performed to analyze
how the finite resolution of the retarding field analyzer affects the measurement, are presented in section
\ref{sec:resolution_GPT}. Section \ref{sec:energy_distribution_measurements} first introduces some methods used to
optimize the outcome of the experiments and then shows energy distribution measurements for different extraction
fields, laser intensities, beam current and beam energy. The conclusions are summarized in section
\ref{sec:energy_spread_conclusion}.

\section{Experimental setup \label{sec:energy_spread_exp_setup}}

The ion beam analyzed in this research is created by two-step photoionization of a laser-cooled and compressed beam of
$^{85}$Rb atoms. The details of the creation and analysis of this atomic beam were described earlier
\cite{tenHaaf2017_atombeam}. There, also a measurement of the longitudinal velocity distribution of the atoms is shown.
From these laser-induced fluorescence results, the full width 50\% energy spread of the  incoming atoms is
determined to be 4 meV. As will be shown this is much smaller than the energy spread caused by the variation in ionization
position.

\begin{figure}[t!]
	\includegraphics{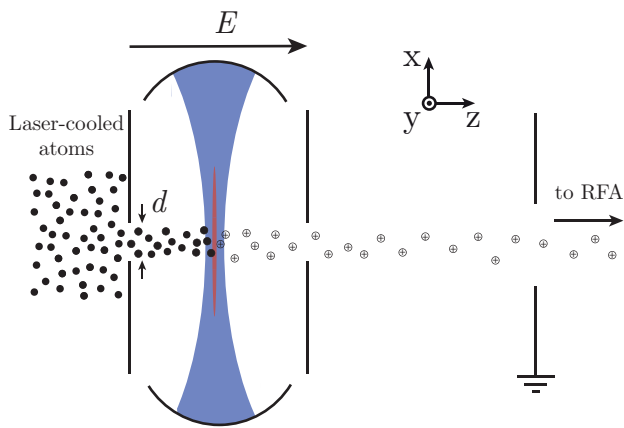}
	\caption{Schematic picture of the photoionization and acceleration setup in the ABLIS setup. The atomic beam is
	skimmed with a selection aperture to only let through the ions which are desired to be ionized. The ionization takes
	place at the crossover of a tightly focused excitation laser beam and an ionization laser beam whose intensity is
	enhanced in a build-up cavity. This crossover is positioned in the first of two acceleration stages in which an
	extraction electric field $E$ is present. Acceleration takes place in two steps in order to set the extraction field
	separately from the final beam energy.}
	\label{fig:photoionization_for_energyspread}
\end{figure}

The photoionization and acceleration setup is schematically depicted in figure
\ref{fig:photoionization_for_energyspread}. After laser cooling and compression the atomic beam is skimmed by an
aperture with diameter $d$ on a movable aperture strip. Photoionization takes place in the crossover of a tightly
cylindrically focused excitation laser beam and an ionization laser beam whose intensity is enhanced in a build-up
cavity. More details of the ionization setup are given in Ref. \cite{tenHaaf2017_current}. Calculations also discussed
in this reference show that for high ionization laser beam intensities the ionization region can be limited in the
z-direction by solely focusing the excitation laser beam very tightly. The $1/e^2$ diameters of the ionization and
excitation laser beams are estimated at 68 $\mu$m and 16 $\mu$m respectively.

Particle tracing simulations that were performed earlier \cite{tenHaaf2014} showed that a reduction of the ion beam
brightness due to disorder-induced heating can be prevented by applying an extraction electric field with such a
magnitude that a so-called pencil beam is created. The extraction field needed to reach the pencil beam regime
increases with the beam current since a larger beam current is made by increasing the size of the selection aperture.
Since the ion beam energy spread is also determined by this field, it will be beneficial to change the extraction field
when the beam current is changed in a future FIB instrument. To enable setting this extraction field $E$ separately from
the total beam energy $U$, the ABLIS accelerator consists of two acceleration sections. The ionization takes place
between two electrodes that near the axis can be approximated by flat plates with an aperture in them. Near the axis,
the separation between the electrodes is 3 mm and the diameter of the aperture in the second electrode through which the ions are
extracted is 2 mm. Further from the axis, the electrodes are separated more to make room for the lens that focuses the
excitation beam. The third electrode is a flat plate with an aperture with a diameter of 7 mm.

Due to the changing electric fields, the acceleration structure causes a lens action on the ion beam, which deflects
ions that are travelling off-axis. This deflection is unwanted as it can decrease the ion beam brightness and, as will
be discussed later, decrease the energy resolution of the retarding field analyzer. Therefore, the ions need to be
created on the axis of the accelerator. To facilitate alignment, the whole accelerator structure is placed on a flexure
hinge that enables planar movement in the transverse directions. The position of the accelerator can be adjusted with
an accuracy of 22 $\mu$m with two micrometer feedthroughs.

\begin{figure}[t!]
	\includegraphics{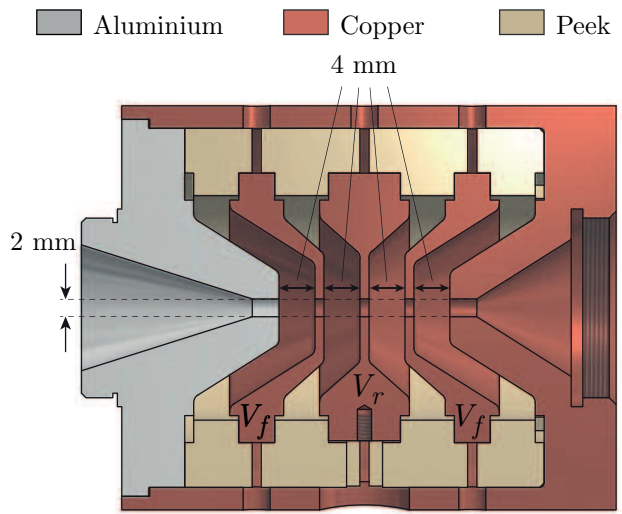}
	\caption{Cross section of the retarding field analyzer (RFA) that is used to measure the energy distribution. The ion
	beam enters the RFA from the left. The  legend indicates what materials the colors correspond to. The three middle
	electrodes are all electrically isolated from the rest. The middle electrode is set at a potential $V_\text{r}$, so
	that only ions with an energy $U>eV_\text{r}$ are transmitted. The second and fourth electrode are set at a potential
	$V_\text{f}$, which is varied in particle tracing simulations in order to maximize the energy resolution. The first
	electrode and the surrounding fifth electrode are grounded. The Faraday cup used to measure the transmitted current is
	placed directly after the RFA, but is not shown in the image. Screw thread on the outside of the extrusion on the left
	is used to connect the RFA to the last electrode of the accelerator.}
	\label{RFA}
\end{figure}

In a retarding field analyzer the incoming ions are decelerated by a retarding potential $V_\text{r}$. If this potential
exceeds the energy of an incoming ion, the ion is reflected. Otherwise the ion is collected by some measurement
device or reaccelerated and collected at a later stage. By varying $V_r$ and measuring the collected or transmitted
current, the energy distribution of the ions is determined. A cross section of the RFA used here is shown in figure
\ref{RFA}. It is similar to the intermediate image lens arrangement described by Simpson \cite{Simpson1961}. In
such an analyzer the ions are decelerated in two stages, giving rise to a lens action with which the ions can be
focused on the center electrode. The top of the potential barrier is positioned in free space within the aperture of
this center electrode. After the retarding electrode the ions are accelerated again to their original energy and
collected by a Faraday cup (not shown in figure \ref{RFA}). The apertures in all five electrodes have a 2 mm diameter.
The middle three electrodes are 1 mm thick in the center. The second and fourth electrode are set at an electric
potential $V_\text{f}$. The RFA is connected to the last accelerator electrode by means of a screw thread on the outside
of the extrusion on the left side in figure \ref{RFA}. This means that the RFA moves simultaneously with the accelerator
when it is translated.


\section{Retarding field analyzer energy resolution\label{sec:resolution_GPT}}

There are two aspects that determine the resolution of the retarding field analyzer. The first one is the variation of
the height of the potential barrier over the position within the aperture in combination with the size of the beam. The
other one is the direction of the local field in combination with the angular spread with which the ion beam approaches
the potential barrier. Both, the size of the beam as well as its angular spread are influenced by the potential on the
focusing electrode. In this section the optimal $V_\text{f}$, that gives the highest energy resolution, is determined
using General Particle Tracer (GPT) simulations \cite{GPT}.


In the GPT simulation, the ions are initialized inside the accelerator and their equations of motion are solved while
taking into account the electric field of the accelerator and the RFA. These fields are interpolated on a cylindrically
symmetric electric field map calculated with CST EM-studio \cite{EMstudio}. Also the geometries are taken into account,
i.e., an ion that does not go through an aperture is removed from the simulation. In this way an RFA measurement can be simulated by
determining the fraction of ions that is transmitted through all apertures as a function of $V_\text{r}$. The full
width 50\% apparent energy spread, i.e., including finite resolution effects, is extracted from the simulation. The
transverse position and velocity distribution are initialized according to the previously introduced Monte Carlo
procedure which takes into account the effect of the selection aperture \cite{tenHaaf2017_current}. Similar as in most
experiments, the diameter of the selection aperture was 21 $\mu$m. The longitudinal position distribution, giving rise to the energy
spread, is calculated from a numerical solution to the optical Bloch equations including ionization
\cite{tenHaaf2017_current}. This distribution with a full width 50\% position spread of 3.3 $\mu$m is calculated for an
ionization laser intensity of $1.7\times10^{10}$ W/m$^2$ and an excitation laser saturation parameter of
$1.3\times10^2$. These values are chosen since they resulted in the lowest energy spread in the experiment, as will be
shown in section \ref{sec:energy_distribution_measurements}.

\begin{figure}[t!]
	\begin{tabular}{l}
			\includegraphics{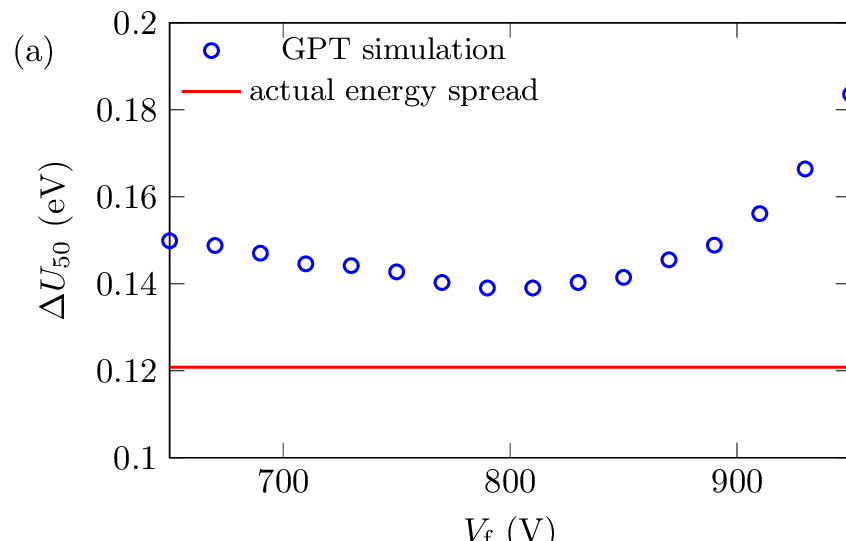}\\
			\includegraphics{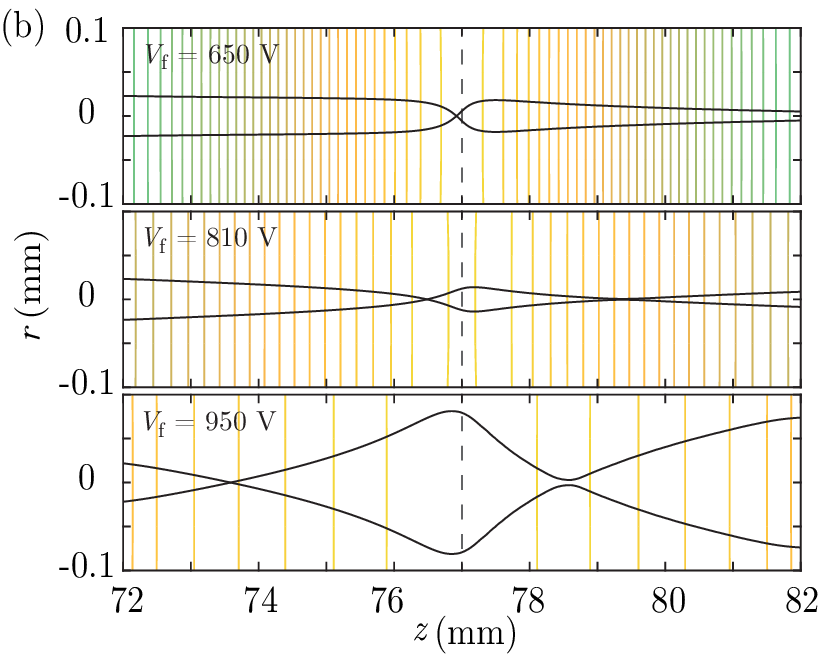}\\
	\end{tabular}
	\caption{Particle tracing simulation results of energy spread measurements using the retarding field analyzer.
	(a) Apparent energy spread of the ion beam, i.e., including finite resolution effects (markers) as a function of the
	focusing voltage $V_\text{f}$. The line indicates the actual energy spread, directly calculated from the ion
	velocities. (b) Beam envelopes containing 50\% of the beam current as a function of the longitudinal position $z$ for
	three different values of the focusing voltage. The retarding electrode is positioned at $z=77$ mm and the focusing
	electrode is positioned at $z=72$ mm. These positions are relative with respect to the position of ionization. The
	simulation were performed for: $U=1$ keV, $E=37$ kV/m and $d=21$ $\mu$m.}
	\label{RFA_resolution}
\end{figure}

The results of the simulations are summarized in figure \ref{RFA_resolution}. In these simulations the ions are
extracted with an extraction field of 37 kV/m and the beam energy was 1 keV. Figure \ref{RFA_resolution}a shows the full
width 50\% energy spread $\Delta U_{50}$ as a function of $V_\text{f}$. The markers show the apparent energy spread
obtained using the RFA, while the line indicates the actual energy spread directly calculated from the incoming ion
velocities. The highest resolution is obtained around $V_\text{f}\approx800$ V, where the overestimation of the energy
spread is only 20 meV. Between $V_\text{f}=750$ V and $V_\text{f}=850$ V the resolution is good and relatively
insensitive for changes $V_\text{f}$. The trend of the RFA resolution can be understood from the beam envelopes plotted
in figure \ref{RFA_resolution}b for three different $V_\text{f}$. For $V_\text{f}=650$ V the beam is focused almost
exactly at the center of the retarding electrode which is indicated by the dashed line. Therefore the analysis suffers
very little from a difference in height of the potential barrier. However, the ions travel towards the center of the
aperture, while the local field is also pointing from the rim of the aperture to its center. A higher resolution is
reached when the beam is slightly divergent when it travels towards the potential barrier, since the ion trajectories
are then better aligned in the direction opposing the local field. This situation is reached for $V_\text{f}=810$ V as
can be seen in the second plot of figure \ref{RFA_resolution}b. When $V_\text{f}$ is increased further the resolution
decreases again because the beam probes a larger region of the retarding electrode aperture, as be seen in the last
plot.


The GPT simulations show that the energy distribution is slightly broadened due the the finite resolution of the RFA.
In the case of a 1 keV beam, the apparent energy spread is only one sixth higher than the expected energy spread of the
ions. However, note that in the simulation a perfect alignment is assumed, i.e., the beam is perfectly aligned with the
RFA axis and the electrodes are also perfectly coaxial. Misalignment decreases the RFA resolution. This is further
discussed in subsection \ref{subsec:vsE}. Furthermore, the smallest resolvable energy spread scales with the beam
energy since the potential drop from the rim of the retarding electrode aperture to the axis becomes larger. This is
discussed in subsection \ref{subsec:vsU}.

\section{Measurements\label{sec:energy_distribution_measurements}}

In this section the performed energy distribution measurements are presented. Subsections
\ref{subsec:Aligment_EnergySpread} and \ref{subsec:Averaging_EnergySpread} introduce the methods to align the
beam and average the results. Subsections \ref{subsec:vsE}-\ref{subsec:vsU} discuss the energy distributions obtained 
for different experimental settings.

\subsection{Alignment\label{subsec:Aligment_EnergySpread}}

\begin{figure*}
	\begin{tabular}{l}
		\includegraphics{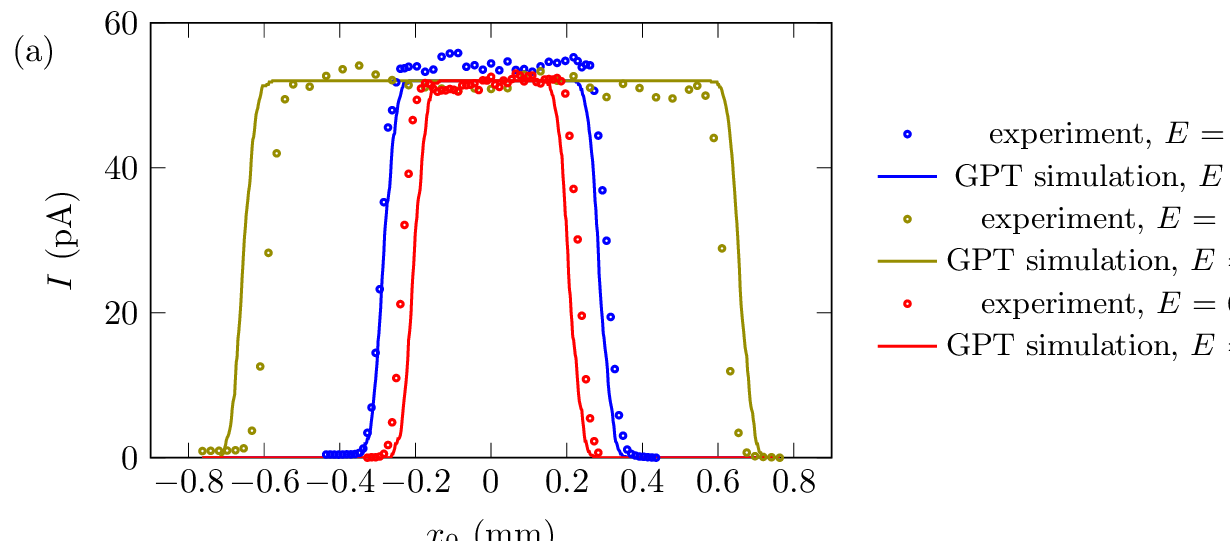}\\
		\includegraphics{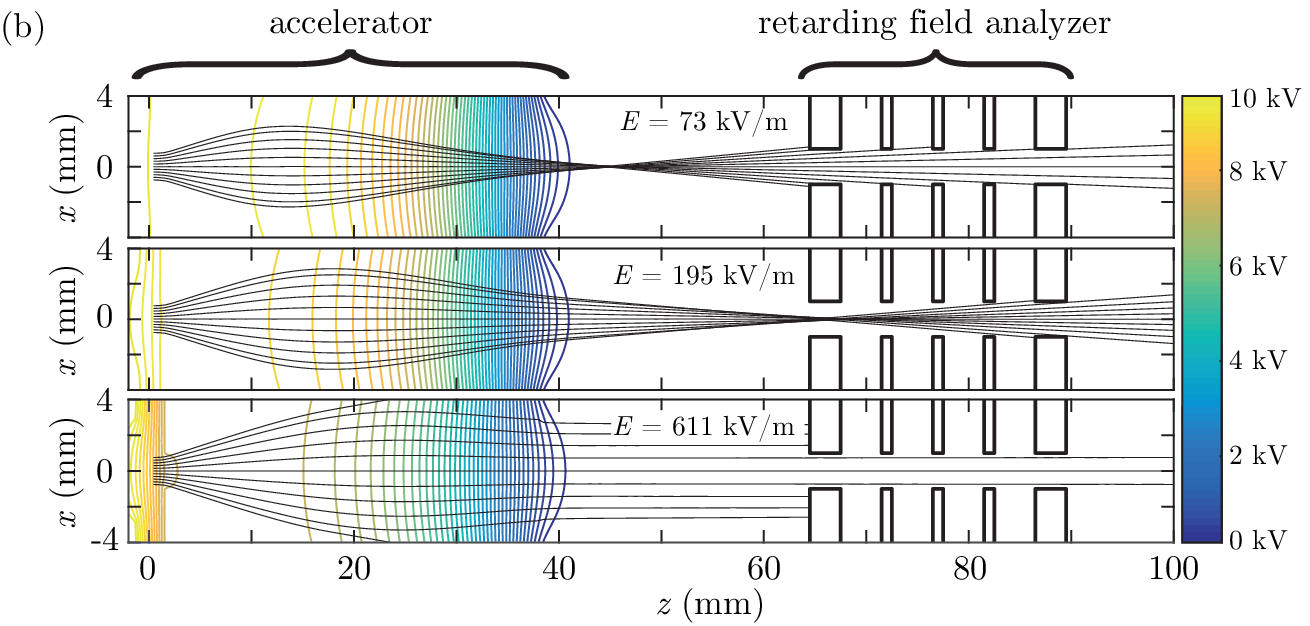}\\
	\end{tabular}
	\caption{(a) Experimental data (markers) and simulation data (lines) of the beam current transmitted through the
	apertures in the retarding field analyzer electrodes as a function of the center transverse position $x_0$ at which
	the ions are created. The measurement was performed with the three values of the extraction field $E$ indicated in the
	legend. (b) Equipotential map of the accelerator for the three extraction fields in (a) and the average ion beam
	trajectories (black lines) for 11 different $x_0$. The thick black lines indicate the contours retarding field analyzer
	electrodes.}
	\label{accelerator_offset}
\end{figure*}

The ions should be created on the axis of the system to obtain the highest energy resolution with the retarding field
analyzer. In order to do so, the current is measured as a function of the accelerator position. In these measurements
there was no voltage applied to any of the RFA electrodes which were only used to cut off the beam when it is deflected
too far from the axis. The result of three of such measurements at different extraction field strengths is shown in
figure \ref{accelerator_offset}. The markers show the measured current as a function of the initial position $x_0$ of
the ion beam with respect to the accelerator axis. The centers of the three flat top profiles indicate the position at
which the position of ionization is exactly on the axis.

Interestingly the flat top profile first increases in width for an increasing extraction field and then decreases. This
can be explained by looking at particle tracing simulation results which are also plotted in figure
\ref{accelerator_offset}a and show the same behaviour. Figure \ref{accelerator_offset}b shows the average trajectories
of the ions in these simulations for several different $x_0$ and for the three different extraction fields. The figure
also shows the equipotential lines of the accelerator and the contours of the RFA electrodes. From these ion beam
trajectories the changes in the width of the profiles in figure \ref{accelerator_offset}a can be understood. Due to the
curvature of the electric field at the end of the extraction region the ions experience an exit kick that diverges the
ion beam, but in this case more important, also bends the whole beam outwards when it is created off-axis. In the
second acceleration stage the deflected ions then first experience a force directed to the axis in the increasing field
section and then again a force directed from the axis in the decreasing field section. As can be seen, this causes the
beam to go through a crossover at low extraction fields. The overall effect of an increase in extraction field and
corresponding decrease in post-acceleration field is that the position of this crossover is displaced to larger $z$,
i.e., the beam leaves the accelerator less converging. From the trajectories in figure \ref{accelerator_offset}b it can
be concluded that when the crossover is inside the RFA the profile in \ref{accelerator_offset}a is the widest because
the displacement from the axis at the position of the RFA electrodes is then smallest for a given $x_0$.

\subsection{Averaging method\label{subsec:Averaging_EnergySpread}}

\begin{figure}[t!]
	\begin{tabular}{l}
			\includegraphics{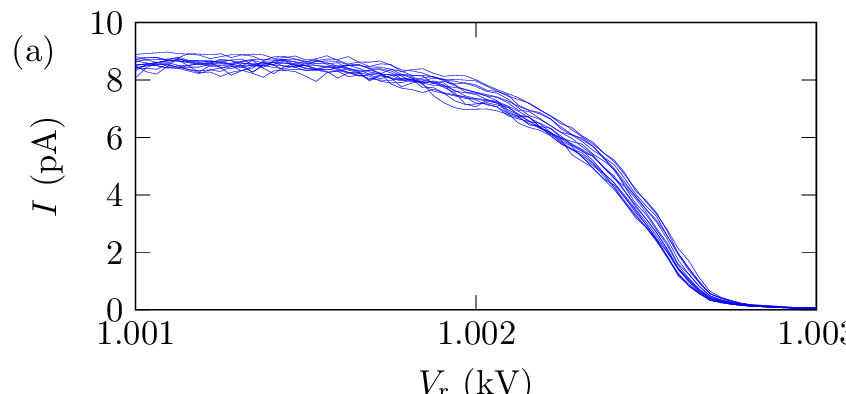}\\
			\includegraphics{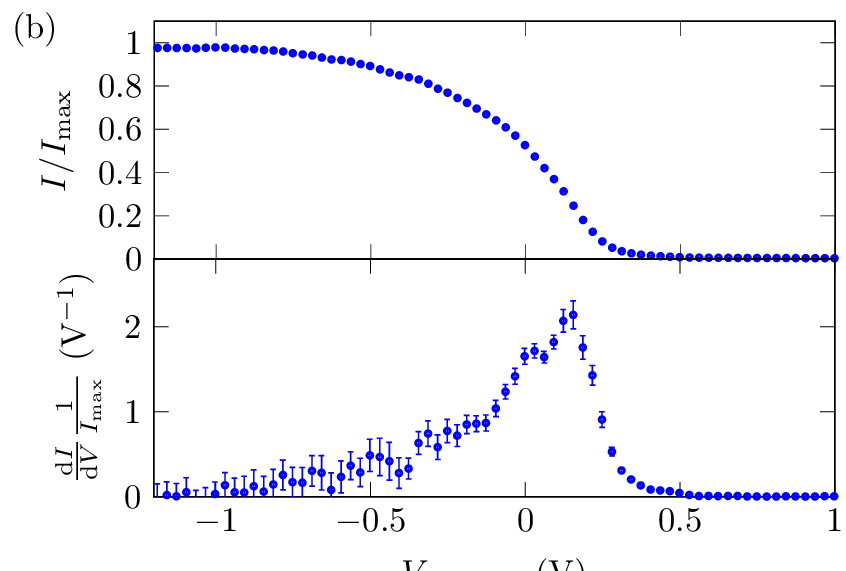}\\
	\end{tabular}
	\caption{(a) Measured current as function of the retarding voltage (15$\times$). (b) The averaged shifted normalized
	current of the curves from (a) (top panel) and its derivative (bottom panel) as a function of the shifted retarding
	voltage. The data is measured with $I_\text{i}=2\times10^{9}$ W/m$^2$, $s=1.3\times10^2$, $U=1$ keV, $E=37$ kV/m,
	$V_\text{f}=850$ V and $d=21$ $\mu$m.}
	\label{method_examples}
\end{figure}

Most of the experimental energy distributions shown in this manuscript are measured by averaging 15 curves of the
current $I$ as a function of the retarding voltage $V_\text{r}$. Figure \ref{method_examples}a shows a typical example
of these 15 unaveraged curves. Measuring a single curve takes approximately two and a half minutes. As can be seen the
curves are all very similar apart from noise, small energy drifts and amplitude drifts. To reduce the noise, especially
in the numerical derivatives, the curves are averaged. However, note that the individual curves are all slightly shifted
with respect to each other, i.e., there is a drift in the beam energy. The reason for this shift is not known, but a
possible explanation could be charging of dielectric surfaces near the accelerator structure. As this drift takes place
over a rather large timescale it is not considered an energy spread. Therefore it is corrected for by shifting the
individual curves with the voltage at which half of the maximum current is reached. This value was found from a linear
fit through the data points between 25\% and 75\% of the maximum current. Furthermore, the curves are divided by the
maximum current $I_\text{max}$, so that the derivative of the curve results in a normalized energy distribution. The
resulting averaged shifted normalized curves are shown in the top panel of figure \ref{method_examples}b. The
derivative of this curve  results in the normalized energy distribution and is shown in the bottom panel of figure
\ref{method_examples}b. The error bars indicate the standard deviation of the mean value.

In the next sections, results are shown of systematic variations of experimental parameters, such as the ionization
intensity $I_\text{i}$, the excitation saturation parameter $s$ and the acceleration field. In order to quantitatively compare
these experimental results with each other a full width 50\% energy spread $\Delta U_{50}$ is determined from the
energy distributions. This value is the smallest width of the distribution that contains 50\% of the normalized current. For
the data shown in figure \ref{method_examples}b, this resulted in $\Delta U_{50}=\left(0.290\pm0.009\right)$ eV, in
which the uncertainty indicates the 68\% confidence interval.

\subsection{Ionization laser beam intensity}

\begin{figure}[t!]
	\begin{tabular}{l}
			\includegraphics{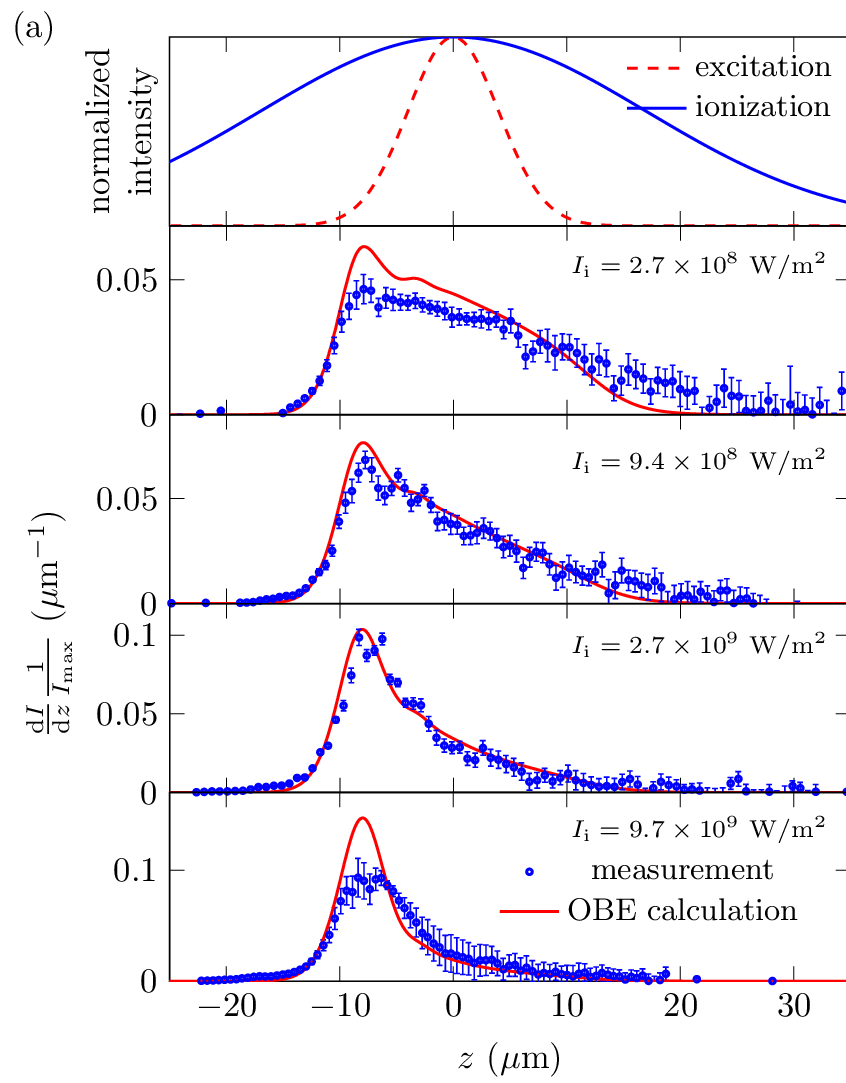}\\
			\includegraphics{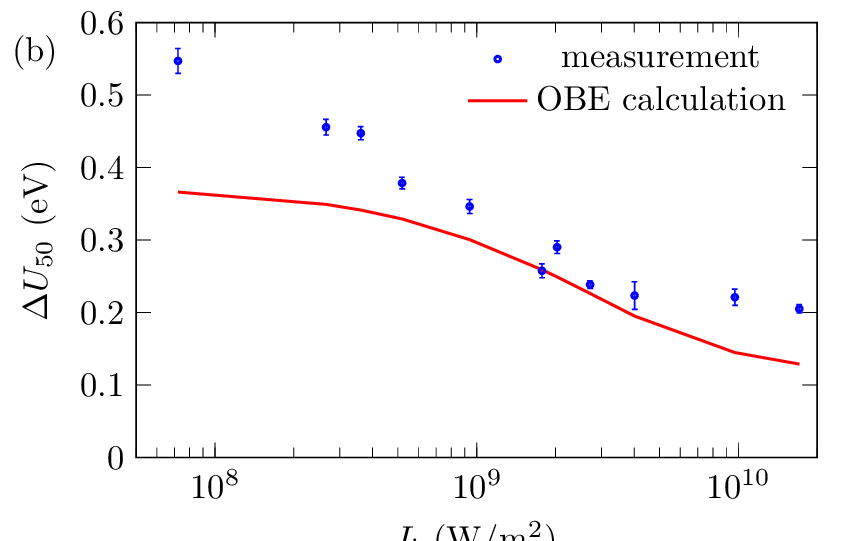}\\
	\end{tabular}
	\caption{(a) Measured (markers with error bars indicating the standard deviation) and calculated (lines) ionization
	position distributions for four different values of the ionization laser beam intensity (bottom four panels). The top
	panel shows the shape of excitation and ionization laser beams that are used in the calculation. (b) Measured (markers
	with error bars indicating the standard deviation) and calculated (lines) full width 50\% energy spread as a function
	of the ionization laser beam intensity. Experiments are performed with $s=1.3\times10^2$, $E=37$ kV/m, $U=1$ keV,
	$V_\text{f}=850$ V and $d=21$ $\mu$m.}
	\label{fig:dU_vs_ion}
\end{figure}

The longitudinal extent of the ionization region in the ABLIS setup (see
figure \ref{fig:photoionization_for_energyspread}), and thus also the energy spread of the resulting ion beam, is
limited by the size of the excitation laser beam. However, the exact distribution is also influenced by the ionization laser
beam intensity $I_\text{i}$. Here the effect of $I_\text{i}$ on the ion energy distribution is determined
experimentally. Since the ion energy is directly proportional to the extraction field, the ionization position $z$ can
be calculated by the transformation,
\begin{equation}
z=-\frac{V_\text{r}}{E}+\zeta
\end{equation}
in which $\zeta$ is a constant determined by the absolute ionization position. With this transformation the ion energy
distribution can be transformed in the ionization position distribution. This allows for comparison with solutions of
the optical Bloch equations including ionization. Figure \ref{fig:dU_vs_ion}a shows the measured ionization position
distributions for four different values of $I_\text{i}$ together with calculated distributions. The constant $\zeta$ is
determined by a fit of the measured distribution with the calculated one for each distribution separately. The position
$z=0$ corresponds to the center of the laser beams as can be seen in the top panel.

The resemblance between the measurement and calculation is remarkable. The measured data reproduces the shape of the
distribution very well. At the negative $z$-side, corresponding to the highest energies, there is a sharp rise in the
distribution due to the onset of excitation. The tail on the positive z-side gets smaller for increasing $I_\text{i}$.
In the distribution for $I_\text{i}=2.7\times10^8$ W/m$^2$ ionization stops because the atoms have returned to the
ground state. For $I_\text{i}=9.7\times10^9$ W/m$^2$ the ionization is quenched since most of the atoms are ionized.
Figure \ref{fig:dU_vs_ion}b shows the resulting energy spread as a function of $I_\text{i}$. The trend of the measured
energy spread and its calculated counterpart is similar. However, at low $I_\text{i}$ the calculation predicts a
somewhat lower energy spread. A possible explanation for this difference can be an experimental excitation laser beam
shape not fully resembling the one in the calculation. The smallest measured energy spread is
$\left(0.205\pm0.006\right)$ eV, corresponding to a full width 50\% ionization position distribution width of
$\left(5.6\pm0.2\right)$ $\mu$m.


\subsection{Excitation laser beam intensity}

\begin{figure}[t!]
	\begin{tabular}{l}
			\includegraphics{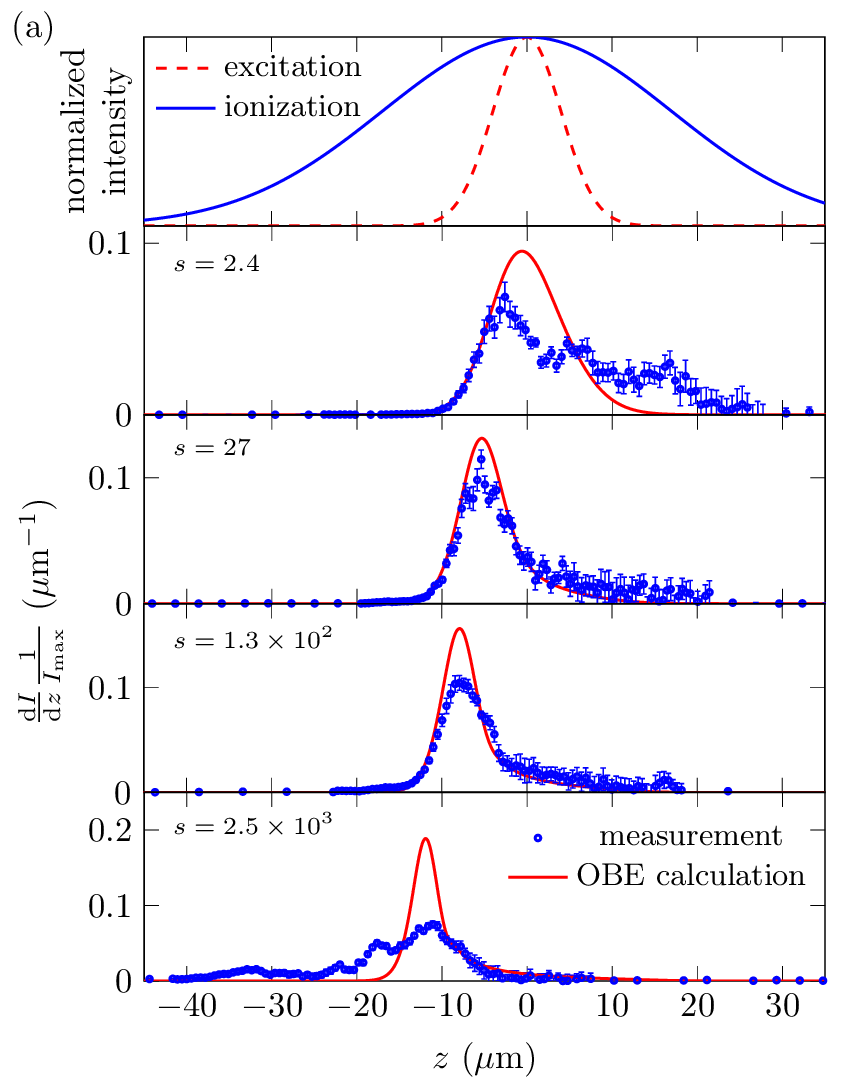}\\
			\includegraphics{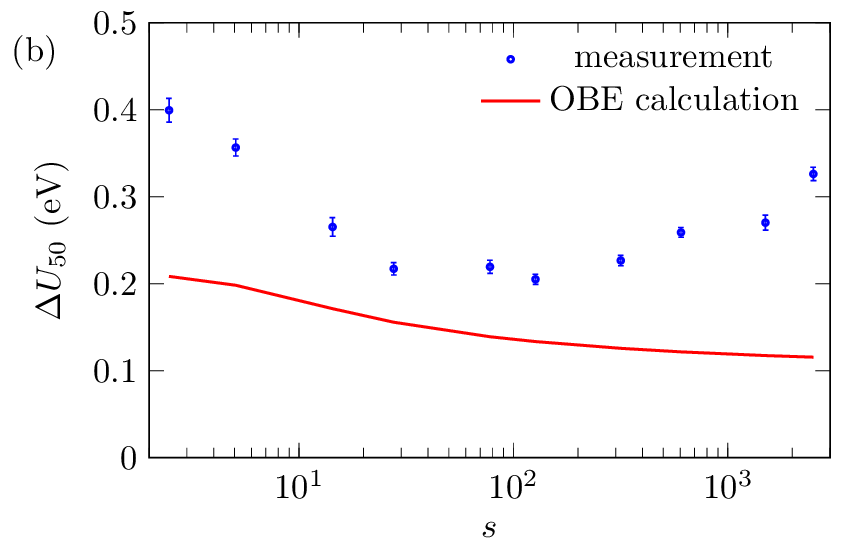}\\
	\end{tabular}
	\caption{(a) Measured (markers with error bars indicating the standard deviation) and calculated (lines) ionization
	position distributions for four different values of the excitation laser beam saturation parameter (bottom four
	panels). The top panel shows the shape of excitation and ionization laser beams that are used in the calculation. (b)
	Measured (markers with error bars indicating the standard deviation) and calculated (lines) full width 50\% energy
	spread as a function of the excitation laser beam saturation parameter. Experiments are performed with
	$I_\text{i}=1.7\times10^{10}$ W/m$^2$, $E=37$ kV/m, $U=1$ keV, $V_\text{f}=850$ V and $d=21$ $\mu$m.}
	\label{fig:dU_vs_exc}
\end{figure}

The ion energy distribution is also measured for varying excitation laser beam intensity $I_\text{e}$, expressed in
terms of the saturation parameter $s=\frac{I_\text{e}}{I_{s}}$. The values of $s$ are determined at the center of the
excitation beam and are calculated with a saturation intensity $I_{s}=51$ W/m$^2$ \cite{tenHaaf2017_current}.
The results are shown in figure \ref{fig:dU_vs_exc}, which again shows four ionization position distributions and a plot of
the $\Delta U_{50}$ as a function of $s$. The results are measured and calculated for the highest $I_\text{i}$ in figure
\ref{fig:dU_vs_ion}.

For $s=27$ and $s=1.3\times10^2$ there is a reasonable agreement between the measured and calculated distribution. As
can be seen in figure \ref{fig:dU_vs_exc}b the widths of the distributions also agrees best in this region. Furthermore,
the trend of a decreasing distribution width is also as expected at low $s$. However, for $s>200$ the width increases
again. The reason for this is most likely that the excitation laser beam shape did not fully resemble a Guassian far
from its center on the negative z-side. On the positive z-side, the laser beam shape did probably not resemble a
Gaussian at all since the measured distribution for $s=2.4$ is much wider on this side. For higher $s$ this does not
result in longer tails on the positive z-side because most of the atoms are already ionized before reaching it. A cause
for such an excitation beam shape can be that it did not go through the focusing lens in its center, which due to
interference and aberrations can lead to an asymmetric shape in the beam waist. All other measurements shown in this
manuscript are measured with $s=1.3\times10^2$.

\subsection{Extraction field strength\label{subsec:vsE}}

\begin{figure}[t!]
	\includegraphics{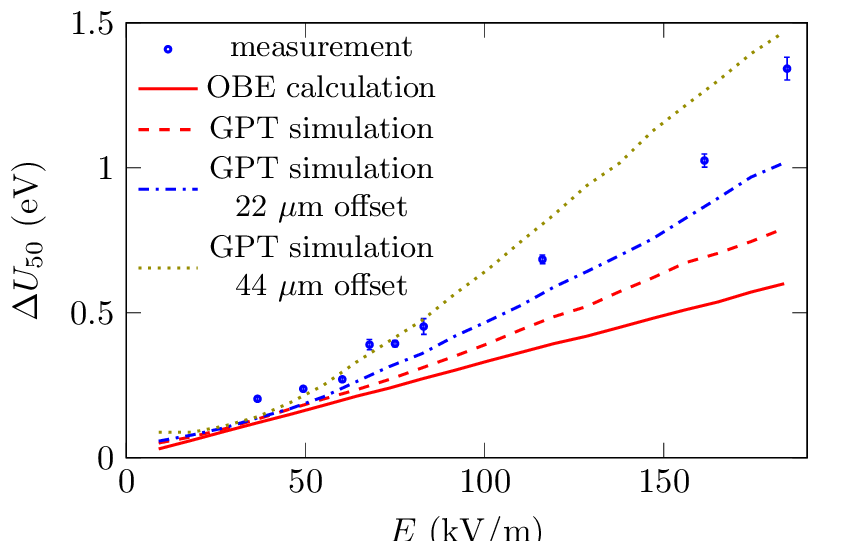}
	\caption{Full width 50\% energy spread as a function of the extraction field. The markers show the
	experimentally determined energy spread and the error bars the standard deviation on this value. The solid line shows
	the expected linear relation between $\Delta U_{50}$ and $E$. The slope of this line is the spread in the ionization
	position which is found from a solution of the optical Bloch equations including ionization. The three dashed/dotted
	lines show the results of particle tracing simulations in which the experiment was simulated for three cases: ions
	created on the axis of the system, ions created at 22 $\mu$m from the axis and ions created at 44 $\mu$m from the
	axis. The experimental results shown are measured with $I_\text{i}=1.7\times10^{10}$ W/m$^2$, $s=1.3\times10^2$, $U=1$
	keV, $V_\text{f}=850$ V and $d=21$ $\mu$m.}
	\label{dE_vs_Field}
\end{figure}

As was explained in section \ref{sec:energy_spread_introduction} the energy spread is expected to be proportional to the
extraction field. A measurement was performed to test this expectation. The results are shown in figure
\ref{dE_vs_Field}, which shows the measured energy spread (markers with error bars) as a function of the extraction
field. Without looking at the other lines in the figure yet the energy spread indeed seems to grow linearly with the
extraction field. However, when taking a closer look the energy spread increases slightly faster. This is especially
visible when comparing the measurement with the expected relation that is also plotted (solid line). This relation is
calculated by multiplication of the ionization position distribution width, found by solving the optical Bloch
equations including ionization \cite{tenHaaf2017_current}, with the applied extraction field. At low extraction field
strengths the energy spread is somewhat higher than expected, but it does increase with the expected slope. There are
some fluctuations that are clearly larger than the error margin. These could be caused by a dependence of the
ionization cross section on the electric field \cite{Freeman1979,Harmin1982}. The deviations between expectation and
measurement are more prominent at increased extraction field strengths.


A possible explanation for the deviation between the expected trend and the measurement are misalignments between
the ionization position, the accelerator electrodes and the RFA electrodes. The effect of misalignments is investigated
using similar particle tracing simulations as in section \ref{RFA_resolution}. Figure \ref{dE_vs_Field} shows the particle
tracing results for three different cases: one in which the ions are created exactly on the axis, one where the ions
are created at an offset of 22 $\mu$m and one in which this offset was 44 $\mu$m. The simulations show that the trend
indeed becomes more non-linear for larger misalignments. The reason for this behaviour is that when the extraction
field is increased the ions leave the accelerator under a larger angle due to which they end up further from the axis
at the retarding electrode. At this transverse position further off-axis the voltage changes faster with the transverse
position. This results in a lower resolution of the RFA due to the finite size of the beam.

The offsets of the ionization position with respect to the axis in the simulations in figure \ref{dE_vs_Field} are not
unrealistic given the accuracy of the accelerator translation. Furthermore, misalignments of the individual electrodes
in the accelerator and RFA can also be of the same order. Therefore it is concluded that such misalignments are likely
the cause of the more than linear growth of energy spread with the extraction field. Regardless of the simulations, the
measured data presented in figure \ref{dE_vs_Field} gives an upper bound for the energy spread of the created ion beam
for various $E$.


\subsection{Beam current}

\begin{figure}[t!]
	\includegraphics{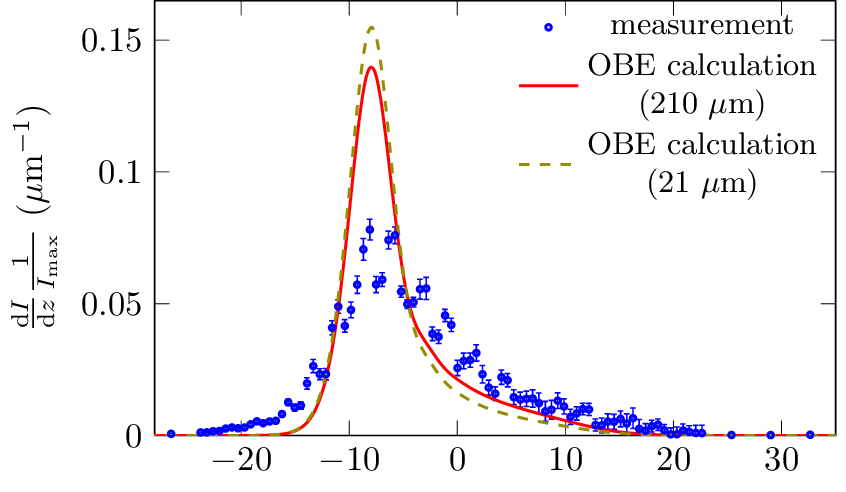}
	\caption{Measured (markers with error bars indicating the standard deviation) and calculated (lines) ionization
	position distribution for a current of 280 pA. The experiment was performed with $I_\text{i}=1.3\times10^{10}$
	W/m$^2$, $s=1.3\times10^2$, $E=37$ kV/m, $U=1$ keV, $V_\text{f}=850$ V and $d=210$ $\mu$m.}
	\label{fig:dU_280pA}
\end{figure}

The measurements shown before this section were all performed on a beam which was selected with a selection aperture
with a diameter of 21 $\mu$m. In this way a beam was created with a current of roughly 9 pA. When selecting the atomic
beam with a larger aperture more current can be created. Figure \ref{fig:dU_280pA} shows the energy distribution for a
current of 280 pA, which was created by using a selection aperture with a diameter of 210 $\mu$m. Note that this
diameter is significantly larger than the $1/e^2$ diameter of the ionization laser beam of 68 $\mu$m. This means that
atoms at a different $y$-position (for orientation reference see figure \ref{fig:photoionization_for_energyspread}) will
experience a different ionization intensity that ranges from $I_\text{i}$ at $y=0$ to almost zero at the edges. These
atoms at the edge will thus ionize over a larger range which increases the energy spread in the beam. As can be seen
from the calculated curves in figure \ref{fig:dU_280pA} this is indeed the case. The calculated ionization position
distribution for a beam selected with a 210 $\mu$m aperture (solid curve) is indeed slightly broader than the
distribution calculated for a selection aperture of 21 $\mu$m (dashed curve). However, the measured distribution is
broader than both of these curves as well as the measured distribution at a lower current under the same circumstances
(see fourth panel in figure \ref{fig:dU_vs_exc}a). A reason for this broadening can be the resolution of the RFA,
especially because the beam is now approximately 3 times larger in the $x$-direction. In the $y$-direction the beam is
also larger, although less due to the limited size of the ionization laser beam waist \cite{tenHaaf2017_current}. Due
to this larger size the beam will probe a larger range of voltages in the aperture of the retarding electrode. However,
note that the measured full width 50\% energy spread of the beam with a current of 280 pA is still only
$\left(0.311\pm0.005\right)$ eV.

\subsection{Beam energy\label{subsec:vsU}}

\begin{figure}[t!]
	\includegraphics{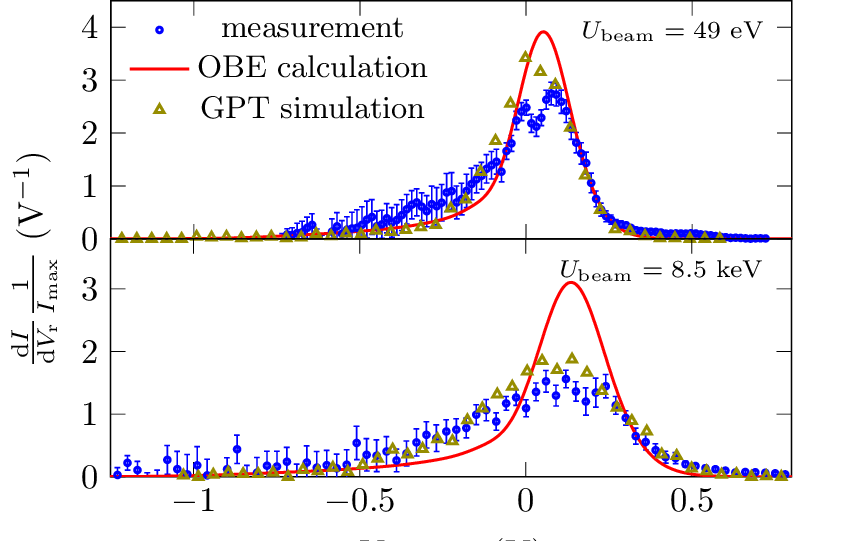}
	\caption{Measured (markers with error bars indicating the standard deviation) and calculated (lines) ion energy
	distribution for ions with an energy of 49 eV (top) and 8.5 keV (bottom). The triangular markers show the result of a
	particle tracing simulation in which the expected distribution of energies is analyzed with the retarding field
	analyzer, thus including finite resolution effects. The experiments are performed with $I_\text{i}=1.9\times10^{10}$
	W/m$^2$, $s=1.3\times10^2$, and $d=21$ $\mu$m.	In the case of $U=49$ eV, $E=41$ kV/m and $V_\text{f}=42$ V. In the
	case of $U=8.5$ keV, $E=51$ kV/m and $V_\text{f}=7375$ V.}
	\label{fig:dU_U}
\end{figure}

The advantage of the two stage acceleration in the ABLIS setup is that a beam with a different energy can be created
while maintaining the same extraction field at the position of ionization. In this way the energy spread of the beam is
independent of the beam energy. The measurements shown before this section were all performed on a beam with an energy
of 1 keV. Figure \ref{fig:dU_U} shows the measured (circular markers with error bars) energy distributions for a beam
with an energy of only 49 eV (top panel) and a beam with an energy of 8.5 keV (bottom panel). Shown are also the
expected distributions from the optical Bloch equation calculation (lines) and simulation results in which the RFA was used to
extract the energy distribution of the beam (triangular markers). Note that the measurements at $U=8.5$ keV are
performed with a slightly higher extraction field of $51$ kV/m as compared to $41$ kV/m for the measurement at $U=49$
eV.

The measured, calculated and simulated energy distributions at $U=49$ eV all agree well with each other. The
measured distribution is slightly broader than the other two, mostly visible by the slightly longer tail on the low
energy side and the higher peak value. The full width 50\% energy spread of the measured distribution amounts
$\left(0.215\pm0.005\right)$ eV. The measured distribution at 8.5 keV differs more from its expected distribution.
However, the simulated distribution which includes finite resolution effects, does agree well with the
measurement. They differ from the calculated distribution because the RFA resolution becomes lower for increasing beam
voltage, since the retarding voltage distribution over the cross section of the aperture in the retarding electrode
scales with the beam energy. The full width 50\% energy spread of the measured distribution amounts
$\left(0.38\pm0.02\right)$ eV. These experiments, together with simulations, show that the energy spread of the beam is
independent of the energy itself and is only determined by the extraction field in which the ions are created and the
ionization position distribution.

\section{Conclusion\label{sec:energy_spread_conclusion}}

The energy distribution of an ultracold ion beam is measured using a retarding field analyzer (RFA). The energy
distribution is measured for different excitation and ionization laser intensities. In general, the distributions show a
large degree of similarity with expected distributions found by numerically solving the optical Bloch equations under
conditions similar as in the experiment \cite{tenHaaf2017_current}. The differences that are present can be explained
by a possibly different excitation laser beam shape in the experiment than in the calculation.

As the ions in the ABLIS setup are accelerated in two steps, it is possible to set the extraction field independent
of the beam energy. Therefore the energy spread is independent of the beam energy, which was verified experimentally
at 49 eV and 8.5 keV. The energy distribution is completely determined by the ionization position distribution,
influenced by the excitation and ionization laser beam intensities, and the extraction electric field. The measured
energy spread increased more than linear with the extraction field. Simulations show that this is likely caused by ions
being created off-axis in the accelerator, which lowers the resolution of the RFA.

The smallest full width 50\% ionization position distribution width measured is $\left(5.6\pm0.2\right)$ $\mu$m. With an
extraction field of 37 kV/m this gives rise to an energy spread of $\left(0.205\pm0.006\right)$ eV. With the
experimentally reached beam density taken into account, this electric field would be large enough to create a pencil
beam containing a current of 3 pA, thus preventing disorder-induced heating \cite{tenHaaf2014}.

\begin{acknowledgments}
This work is part of the research programme Ultracold FIB with project number 12199, which is (partly) financed
by the applied and engineering sciences division of the Netherlands Organisation for Scientific Research (NWO). The
research is also supported by Thermo Fisher Scientific, Pulsar Physics and Coherent Inc.
\end{acknowledgments}

\bibliography{energy_spread}

\end{document}